\documentclass[12pt]{iopart}
\usepackage{iopams,setstack}
\usepackage{graphicx,color}
\newcommand{\EQ }[1]{equation\ (\ref{#1})}
\newcommand{\EQS}[2]{equations\ (\ref{#1}) and (\ref{#2})}
\newcommand{\EQM}[3]{equations\ (\ref{#1}), (\ref{#2}) and (\ref{#3})}
\newcommand{\FIG}[1]{figure\ \ref{#1}}
\newcommand{\SEC}[1]{section\ \ref{#1}}
\newcommand{\SES}[2]{sections\ \ref{#1} and \ref{#2}}
\newcommand{\GES}[3]{#1\ (\ref{#2}) and (\ref{#3})}
\newcommand{\AV}[3]{\langle#1\rangle_{#2}^{#3}}
\def\cH{{\cal H}}
\def\cL{{\cal L}}
\def\cM{{\cal M}}
\def\cV{{\cal V}}
\def\cW{{\cal W}}
\def\cA{{\cal A}}
\def\cB{{\cal B}}
\def\cC{{\cal C}}
\def\cD{{\cal D}}

\def\inverseTL{1.51}
\def\inverseTH{1.05}

\def\figureI{
 \begin{figure}[t]
  \begin{center}
   \includegraphics[width=0.50\linewidth]{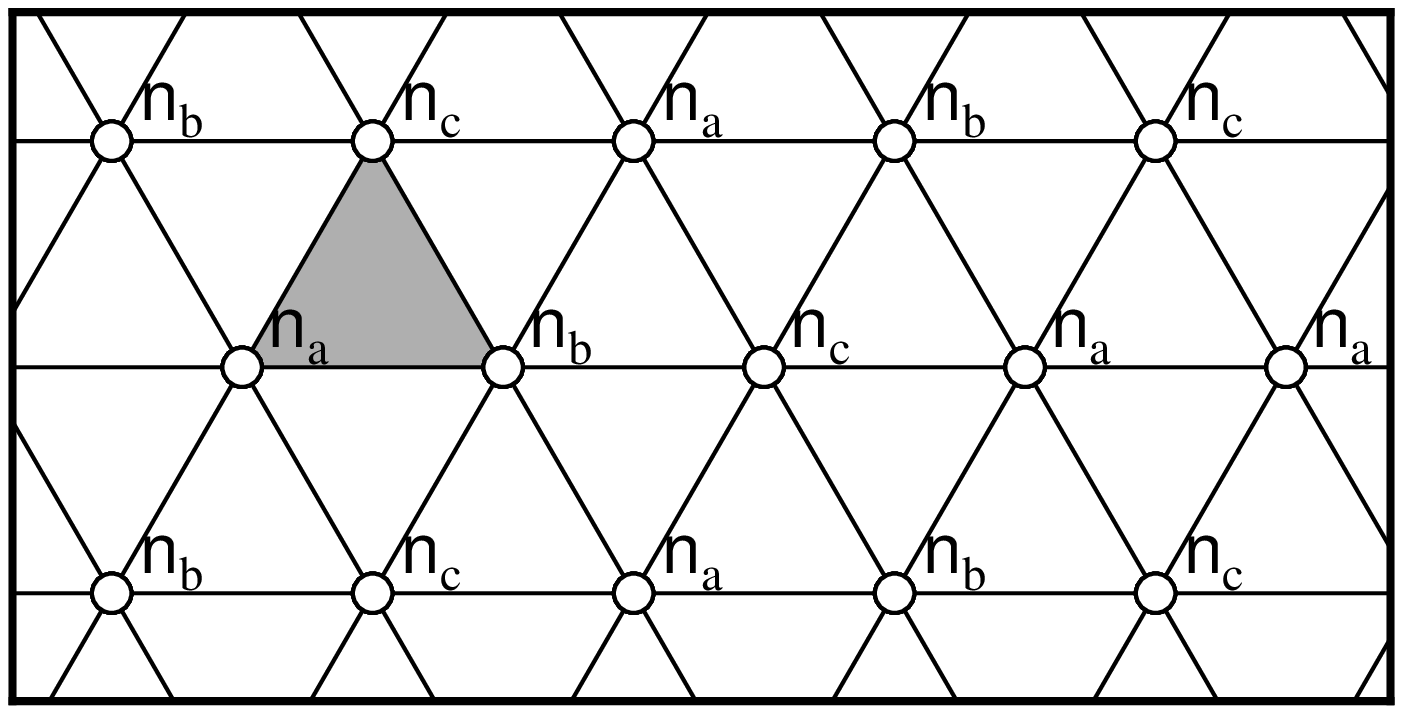}
  \end{center}
  \caption{
  The triangular lattice $\Lambda$ and the sublattice dependent numbers
  $(n_{\rm a},n_{\rm b},n_{\rm c})$ (see the text).
  The shaded area exhibits the elementary plaquette of $\Lambda$.}
  \label{FIG1}
 \end{figure}
}

\def\figureII{
 \begin{figure}[t]
  \begin{center}
   \includegraphics[width=0.60\linewidth]{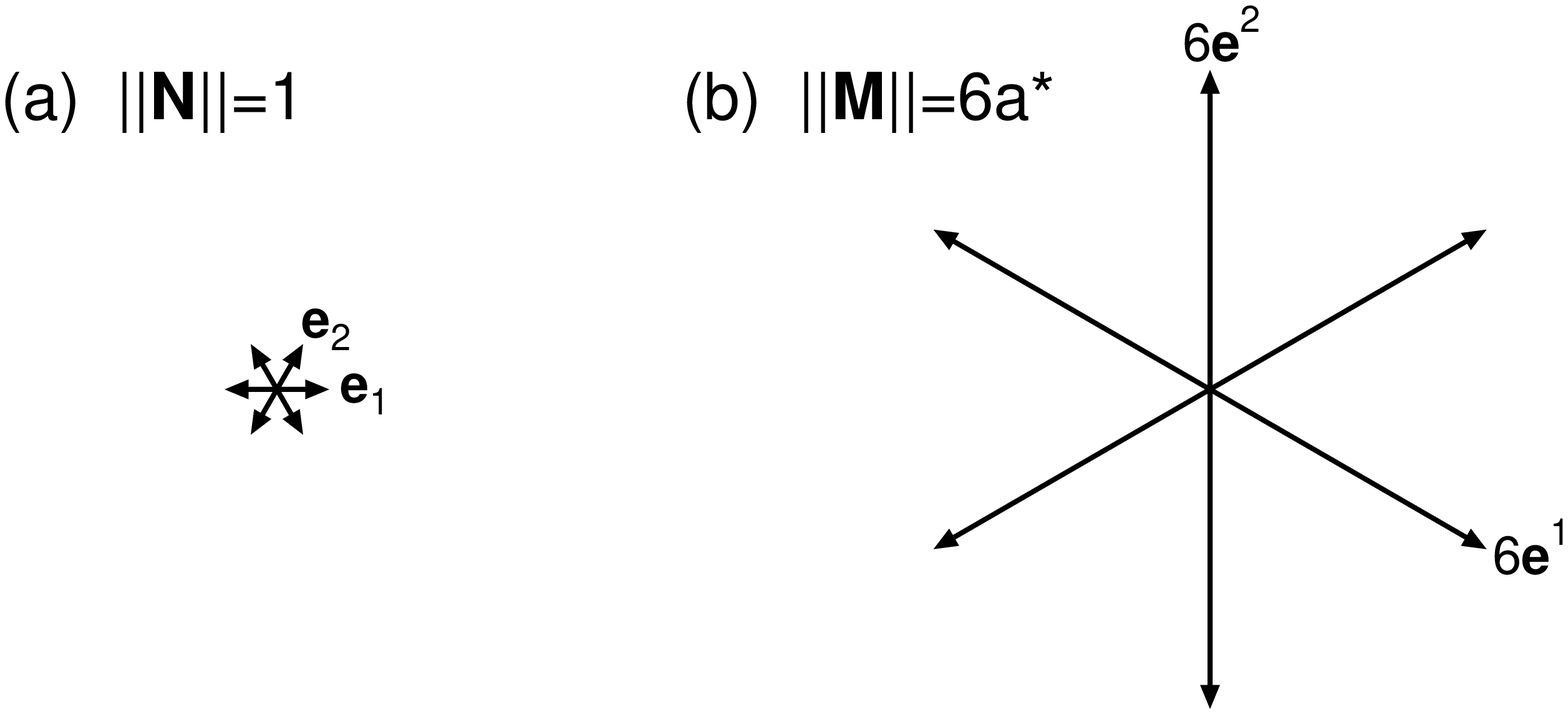}
  \end{center}
  \caption{
  The schematic representation of the vector charges in the Lagrangian
  density $\cL$:
  (a) The magnetic vector charges in $\cL_2$ [\EQ{eq_N}] which represent
  the discontinuities of ${\bf \Phi}$.
  (b) The electric vector charges in $\cL_1$ [\EQ{eq_M}] (the $p=6$ case)
  which bring about the phase locking potential with the $p^2$ minima.}
  \label{FIG2}
 \end{figure}
}

\def\figureIII{
\begin{figure}[t]
 \begin{center}
  \includegraphics[width=0.70\linewidth]{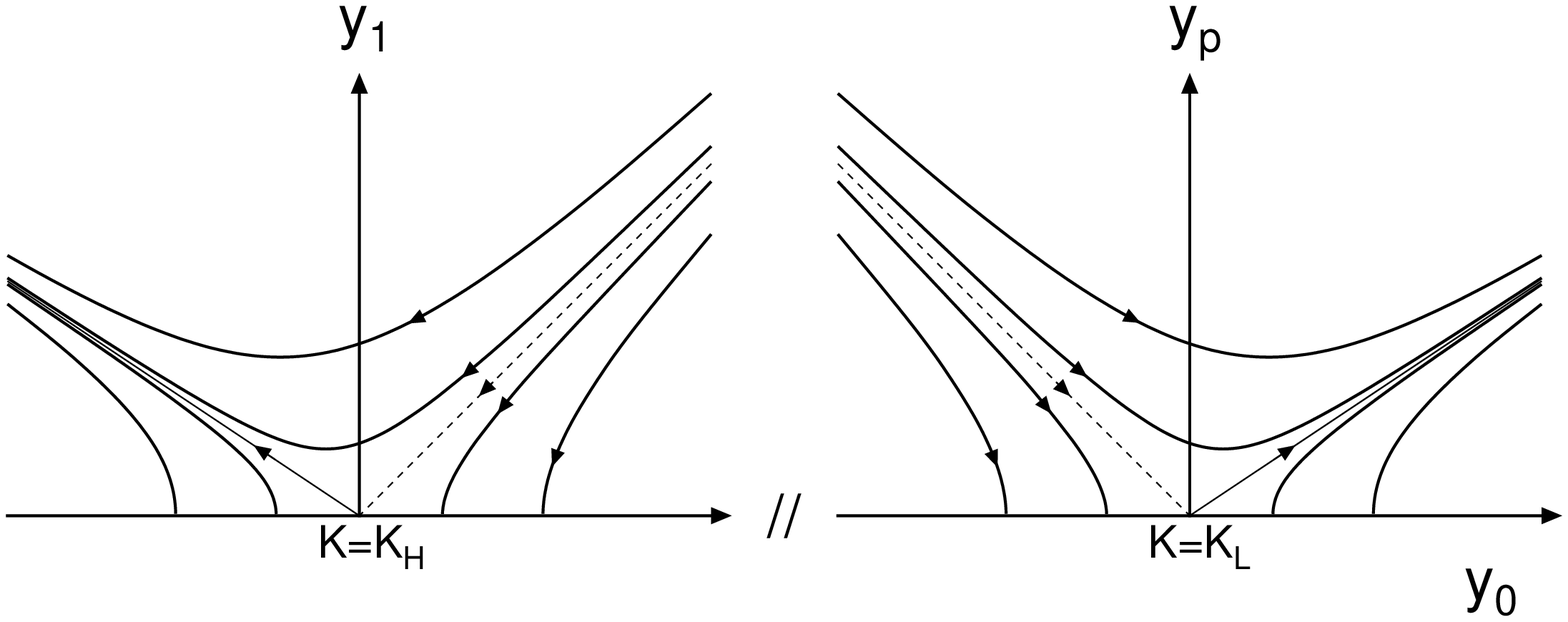}
 \end{center}
 \caption{
 The schematic RG-flow diagram.
 The left (right) panel exhibits the flow around the multicritical
 fixed point $(y_0,y_1)=(0,0)$  [$(y_0,y_p)=(0,0)$] corresponding to
 the transition temperature $T_{\rm H}$ $(T_{\rm L})$.
 The scale of $y_0$ in the left is different from that in the right,
 and the $y_1$ and the $y_p$ axes are not on the same plain.
 The separatrixes around the points are given by the dotted lines.}
 \label{FIG3}
\end{figure}
}

\def\figureIV{
\begin{figure}[t]
 \begin{center}
  \includegraphics[width=0.95\linewidth]{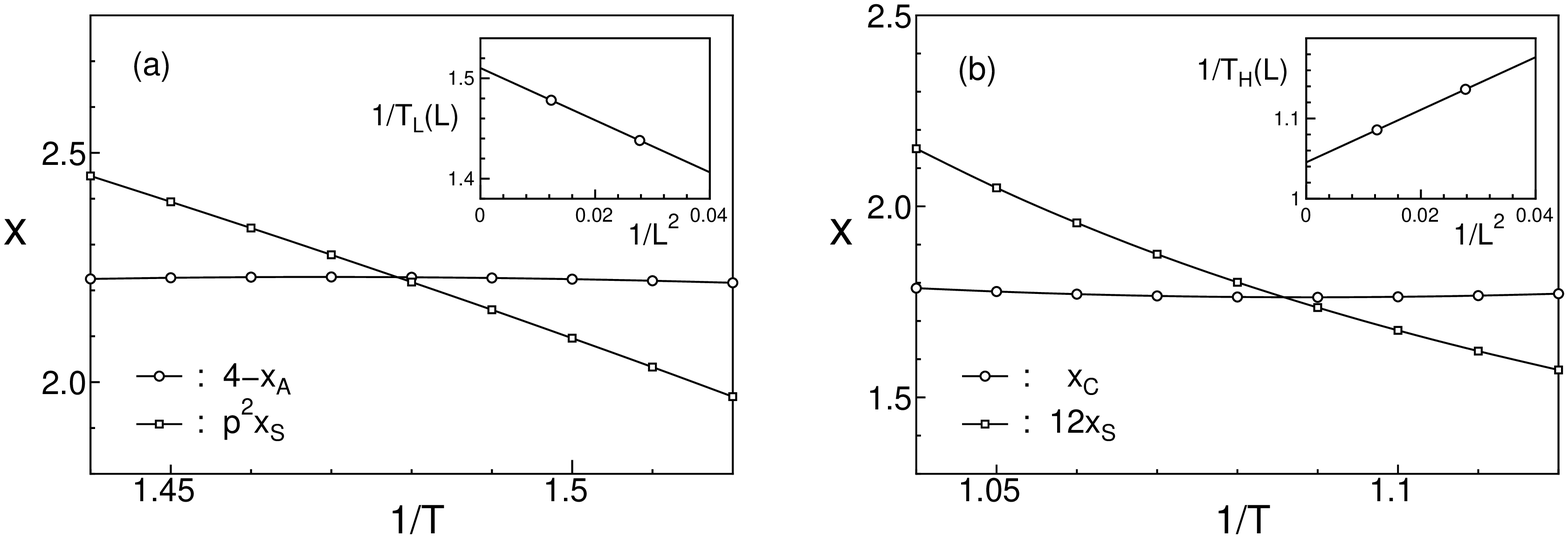}
 \end{center}
 \caption{
 (a)
 The plot of the level-crossing condition of \EQ{eq_LC_TL}
 (the $p=6$ case).
 The circles (squares) with the fitting curve exhibit its rhs (lhs).
 The crossing point gives the finite-size estimate of the phase
 transition point $1/T_{\rm L}(L)$ with $L=9$. 
 The inset shows the extrapolation of the data to the thermodynamic
 limit, and gives $1/T_{\rm L}\simeq\inverseTL$.
 (b)
 The same plot of \EQ{eq_LC_TH} as the panel (a), where the circles
 (squares) with the fitting curve exhibit the rhs (lhs).
 The crossing point estimates $1/T_{\rm H}(L)$ with $L=9$.
 The inset shows the extrapolation of the finite system data, and gives
 $1/T_{\rm H}\simeq\inverseTH$.}
 \label{FIG4}
\end{figure}
}

\def\figureV{
\begin{figure}[t]
 \begin{center}
  \includegraphics[width=0.95\linewidth]{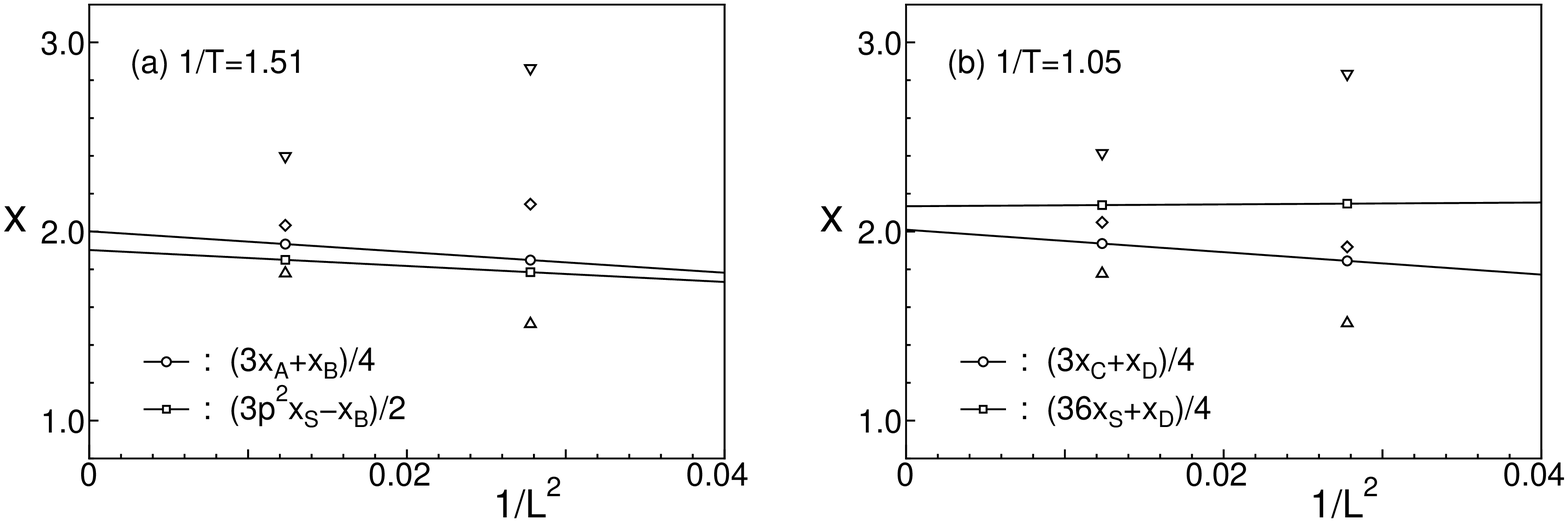}
 \end{center}
 \caption{
 (a)
 The check of the universal relations among scaling dimensions at
 $T_{\rm L}$.
 The circles (squares) with the fitting line plot the lhs of
 \EQ{eq_TL_av}
 [the difference $(3p^2x_S-x_{\cB})/2$] at $1/T=1.51$.
 The up- and the down-ward triangles show $x_{\cA}$ and $x_{\cB}$,
 respectively, and the diamonds plot $p^2x_S$ (with $p=6$).
 (b)
 The check of the universal relations at $T_{\rm H}$.
 The circles (squares) with the fitting line plot the lhs of
 \EQ{eq_TH_av}
 [the average $(36x_S+x_{\cD})/4$] at $1/T=1.05$.
 The up- and the down-ward triangles show $x_{\cC}$ and $x_{\cD}$,
 respectively, and the diamonds plot $36x_S$.}
 \label{FIG5}
\end{figure}
}

\begin{document}

 \title[Criticality in triangular-lattice three-spin interaction model]
 {Critical intermediate phase and phase transitions in\\
 a triangular-lattice three-spin interaction model:\\
 Level-spectroscopy approach}

 \author{Hiromi Otsuka$^1$ and Kiyohide Nomura$^2$}

 \address{
 $^1$Department of Physics, Tokyo Metropolitan University, Tokyo 192-0397, Japan}
 \address{
 $^2$Department of Physics, Kyushu University, Fukuoka 812-8581, Japan}

 \ead{otsuka@phys.metro-u.ac.jp}

 \begin{abstract}

  \noindent
  We investigate infinite-order phase transitions like the
  Berezinskii-Kosterlitz-Thouless transition observed in a
  triangular-lattice three-spin interaction model.
  Based on  a field theoretical description and the
  operator-production-expansion technique, we perform the
  renormalization-group analysis, and then clarify properties of
  marginal operators near the phase transition points.
  The results are utilized to establish criteria to determine the
  transition points and some universal relations among excitation levels
  to characterize the transitions.
  We verify these predictions via the numerical analysis on eigenvalue
  structures of the transfer matrix.
  Also, we discuss an enhancement of symmetry at the end points of a
  critical intermediate phase in connection with a transition observed
  in the ground state of the bilinear-biquadratic spin-1 chain.
 \end{abstract}
  

 \section{Introduction}

 Phase transitions and critical phenomena observed in classical spin
 systems have been investigated for a long time.
 Their theoretical treatments including numerical ones have revealed a
 variety of features, and also have been offering the interfaces to
 understand real materials.
 At the same time, the universality observed in phase transitions is one
 of the most important concepts. 
 For the two-dimensional (2D) critical systems, it is pronouncedly
 expressed in terms of the conformal symmetry being possessed by
 relevant effective field theories.
 The central charge $c$ is the widely-known theoretical parameter
 \cite{BPZ}.
 In the case $c<1$, it appears to almost characterize a universality
 class, i.e., a possible set of critical exponents
 \cite{FQS}: 
 The discrete systems such as the Ising and the three-state Potts
 ferromagnets show the second-order transitions whose universality
 classes are given by the conformal symmetries with the rational values
 of $c$.
 On the other hand, in the case $c\ge1$, there still exist considerable
 efforts to understand the universalities of phase transitions.

 It is widely known that the systems with strong frustrations sometimes
 exhibit the residual entropies and the critical ground states with
 $c\ge1$.
 Those of the triangular-lattice Ising
 \cite{Wann50,Hout50,Husi50,Step70}
 and the square-lattice three-state Potts antiferromagnets
 \cite{Lena67,Baxt70triangle}
 are the typical ones with $c=1$.
 Further, the Kagom\'e-lattice three-state Potts
 \cite{Baxt70kagome,Huse92}
 and the square-lattice four-state Potts vertex antiferromagnets 
 \cite{Lieb72,Read}
 were clarified to possess the ground states with $c=2$ and $c=3$,
 respectively
 \cite{Kond95,Kond96}.
 In addition, the dimer, the loop-gas, and the coloring models
 (some of them can be related to the spin models) are other examples
 to show the criticality with $c\ge1$
 \cite{Jaco07}.
 Also, the frustration effects can increase the central charge of the
 finite-temperature criticality as observed in the 2D fully frustrated
 XY model
 \cite{Thij90}. 
 So they have been gathering the great attentions for long time both
 theoretically and experimentally.

 On another front, multispin interactions appear to include an effect to
 enhance the central charge:
 The exactly solved Baxter-Wu model consisting of the three Ising-spin
 product interaction is the most basic one
 \cite{Baxt73};
 it shows the second-order transition whose universality is the same as
 that of the four-state Potts ferromagnet
 \cite{Alca99}.
 The Ising and the four-state Potts criticalities are of $c=1/2$ and $c=1$,
 respectively.
 Therefore, the multispin interactions are expected as an another source
 to bring about larger values of $c$, although they have not been argued
 frequently in this context.

 \figureI

 In this paper, we investigate the three-spin interaction model (TSIM)
 introduced a long time ago by Alcaraz {\it et al}
 \cite{Alca82,Alca83}.
 Suppose that $\langle k,l,m\rangle$ denotes three sites at the corners
 of each elementary plaquette (see the shaded area in \FIG{FIG1}) of the
 triangular lattice $\Lambda$ consisting of interpenetrating three
 sublattices $\Lambda_{\rm a}$, $\Lambda_{\rm b}$ and $\Lambda_{\rm c}$,
 then the following reduced Hamiltonian expresses a class of TSIM:
 \begin{equation}
  \cH
   =-\frac{J}{k_{\rm B}T}\sum_{\langle k,l,m\rangle}
   \cos\left(\varphi_k+\varphi_l+\varphi_m\right). 
   \label{eq_Hamil}
 \end{equation}
 One model parameter, the temperature $T$, will be measured in units of
 $J/k_{\rm B}$.
 The angle variables $\varphi_k=2\pi n_k/p$ $(n_k\in[0,p-1])$ are
 located on sites, and define the ${\mathbb Z}_p$ clock variables
 \cite{Jose77}
 (the Baxter-Wu model is contained as its special case of $p=2$).
 As we shall take a quick look at them in \SEC{sec_THEORY},
 the most intriguing ones are the properties of an critical intermediate
 phase theoretically expected for $p\ge4$, and its instabilities to the
 ordered and the disordered phases.
 Alcaraz {\it et al} derived the vector Coulomb gas (CG) representation of
 the model
 \cite{Alca83}. 
 Especially, they provided the renormalization-group (RG) analysis based
 on its similarity to the triangular-lattice defect-mediated melting
 phenomenon which is known as the
 Kosterlitz-Thouless-Halperin-Nelson-Young (KTHNY) theory
 \cite{Kost73,Halp78,Nels78,Youn79,Nels79}.
 In the previous paper, we also argued the effective description of TSIM
 based on the symmetry properties and the so-called ideal-state graph
 concept by Kondev and Henley
 \cite{Kond95,Kond96,Jaco04};
 and then introduced the vector dual sine-Gordon Lagrangian density
 \cite{Otsu07}.
 Since its criticality is of $c$ equal to the number of components of
 the vector field, $c=2$ was theoretically expected.
 We performed the numerical calculations to confirm this and further the
 properties of the low-energy excitations. 
 However, a detailed analysis of the model on and around the transition
 points has not been done yet.
 Here, based on the effective field theory, we shall first perform the
 RG analysis, and derive the RG equations to describe the transitions to
 the ordered and the disordered phases.
 In both cases, we discuss a mixing of marginal operators along a
 separatrix embedded in the RG-flow diagram because the same argument
 was done for the sine-Gordon model
 \cite{Nomu95},
 and its importance has been widely recognized in the discussions of the
 Berezinskii-Kosterlitz-Thouless (BKT) transitions
 \cite{Kost73,Bere71,Kost74}
 (for applications to classical systems, see
 \cite{Mats05,Otsu05a,Otsu06a,Otsu06b}).
 We then clarify excitation spectra characteristic to the phase
 transitions observed.
 For these purposes, we shall utilize some formulae which require the
 so-called conformal field theory (CFT) data such as the scaling
 dimensions and the operator-product-expansion (OPE) coefficients.
 Therefore, we provide detailed explanations of the OPE calculations
 among local density operators in our field theory
 \cite{Polchinski}.

 The organization of this paper is as follows. 
 In \SEC{sec_THEORY}, according to our previous research, we shall
 explain our Lagrangian density to effectively describe the low-energy
 and the long-distance behaviors of TSIM.
 The calculations of OPE coefficients necessary for the CFT technology,
 the RG analysis of phase transitions, and the conformal perturbation
 calculations of excitation spectra up to the one-loop order are
 performed there.
 In \SEC{sec_NUMERICAL}, based on the analysis in \SEC{sec_THEORY}, we
 shall explain our numerical calculation procedure
 (the so-called level-crossing conditions)
 to determine the transition points.
 We perform numerical diagonalization calculations of the transfer
 matrix, and then provide their estimates.
 Further, to serve a reliability, we check some universal relations
 among excitation levels observed in finite-size systems---in short, we
 shall perform the level-spectroscopy analysis of TSIM
 \cite{Nomu95}.
 The \SEC{sec_DISCUSSION} is devoted to discussions and summary of the
 present investigation.
 An enhancement of symmetry at the transition points will be pointed
 out; we shall discuss its connection with a 1D quantum spin system.
 For readers' convenience, we shall provide two appendices:
 In appendix A, we summarize the properties of the critical fixed point
 of our model which include the conformal invariance as well as OPE's
 among basic operators.
 In appendix B, based on appendix A, we provide details in the
 derivations of some useful relations; these will contribute directly to
 the analysis of the critical phenomena observed in the present model.

 \section{Theory} \label{sec_THEORY}

 \subsection{Vector dual sine-Gordon model}

 Since the symmetry property is the key to understand the criticality
 and the phase transitions, we shall begin with its description. 
 Adding to the translations and the space inversions, the model is
 invariant under the global spin rotations
 \begin{equation}
  \varphi_k
  \to
  \varphi_k
  +
  \sum_{\rho={\rm a,b,c}}
  \sum_{l\in\Lambda_\rho}\frac{2\pi n_\rho}{p}
  \delta_{k,l}
 \end{equation}
 with sublattice dependent integers (see \FIG{FIG1}) satisfying the
 condition
 $n_{\rm a}+n_{\rm b}+n_{\rm c}=0$ (mod $p$)
 \cite{Alca83}.
 This symmetry operation---we denote as
 $(n_{\rm a}, n_{\rm b}, n_{\rm c})$---can
 be generated from two of the following three fundamental operations:
 \begin{equation}
  \hat R_{\rm a}: (1,p-1,0),~~~
  \hat R_{\rm b}: (0,1,p-1),{\rm~~~and~~~}
  \hat R_{\rm c}: (p-1,0,1).
  \label{eq_gener}
 \end{equation}
 Thus, it is referred to as the ${\mathbb Z}_p\times{\mathbb Z}_p$ symmetry.
 They satisfy some important relations, e.g.,
 $\hat R_{\rm a}^p=\hat R_{\rm c}\hat R_{\rm b}\hat R_{\rm a}=\hat1$.
 Based on these properties, we introduced the vector dual sine-Gordon
 model in the 2D Euclidean space
 \cite{Otsu07}.
 Writing the Cartesian components of the position vector ${\bf x}$ in
 the space as $(x,y)$ (see \FIG{FIG1}), then it is defined by the
 Lagrangian density $\cL=\cL_0+\cL_1+\cL_2$ with
 \begin{eqnarray}
  &\cL_0
  =
  \frac{K}{4\pi}
  \sum_{i=x,y}
  \|\partial_i{\bf\Phi}({\bf x})\|^2,
  \label{eq_L0}\\
  &\cL_1
  =
  \frac{y_p}{2\pi a^2}
  \sum_{\|{\bf M}\|=pa^*}:{\rm e^{i{\bf M\cdot\Phi({\bf x})}}}: ,
  \label{eq_L1}\\
  &\cL_2
  =
  \frac{y_1}{2\pi a^2}
  \sum_{~\|{\bf N}\|=1~~}:{\rm e^{i{\bf N\cdot\Theta({\bf x})}}}: .
  \label{eq_L2}
 \end{eqnarray}
 The symbol ``: :'' denotes the normal ordering and means the
 subtraction of possible contractions of fields between them.
 We shall employ the same definitions of the fields and the vector
 charges as those in our previous paper:
 ${\bf\Theta}$ is the dual field to ${\bf\Phi}$ and is related as
 ${\rm i}K\partial_i{\bf \Phi}=\epsilon_{ij}\partial_j\bf \Theta$
 ($\epsilon_{ij}$ is the antisymmetric tensor).
 In figure\ 2 of
 \cite{Otsu07},
 we explained the so-called repeat lattice ${\cal R}$ representing the
 periodicity of ${\bf \Phi}$
 \cite{Kond95,Kond96}.
 Using its frame as the Cartesian coordinate, the primitive vectors of
 ${\cal R}$ are given by
 \begin{equation}
  {\bf e}_1=\left(1,0\right){\rm~~~and~~~}
  {\bf e}_2=\Bigl(\frac12,\frac{\sqrt3}{2}\Bigr).
  \label{eq_e^alpha}
 \end{equation}
 Also, the primitive vectors of the reciprocal lattice ${\cal R}^*$ are
 given by
 \begin{equation}
  {\bf e}^1=\Bigl(1,\frac{-1}{\sqrt3}\Bigr){\rm~~~and~~~}
  {\bf e}^2=\Bigl(0,\frac{ 2}{\sqrt3}\Bigr).
  \label{eq_e_alpha}
 \end{equation}
 The magnetic (electric) vector charge ${\bf N}$ (${\bf M}$) is
 quantized in ${\cal R}$ (${\cal R}^*$) whose contravariant (covariant)
 element is expressed as
  $n^\alpha\equiv {\bf e}^\alpha\cdot{\bf N}$
 ($m_\alpha\equiv {\bf e}_\alpha\cdot{\bf M}$),
 and satisfies the condition
 $n^\alpha$ ($m_\alpha$) $\in{\mathbb Z}$.
 Using these vectors, the periodicities of the fields are given by
 ${\bf \Phi}\equiv{\bf \Phi}+2\pi{\bf N}$
 and
 ${\bf \Theta}\equiv{\bf \Theta}+2\pi{\bf M}$.
 The metric tensors for ${\cal R}$ and ${\cal R}^*$ are defined by
 $g_{\alpha\beta}={\bf e}_\alpha\cdot{\bf e}_\beta$
 and
 $g^{\alpha\beta}={\bf e}^\alpha\cdot{\bf e}^\beta$, respectively.
 They satisfy the condition
 $g_{\alpha\gamma}g^{\gamma\beta}=\delta_\alpha^\beta$
 because of the duality relation
 ${\bf e}_\alpha\cdot{\bf e}^\beta=\delta_\alpha^\beta$.
 A small calculation using \EQS{eq_e^alpha}{eq_e_alpha} yields their
 explicit forms:
 \begin{equation}
  g_{\alpha\beta}
   =
   \left(
    \begin{array}{cc}
     1       & \frac12 \\
     \frac12 & 1       \\
    \end{array}
	 \right)
   {\rm~~~and~~~}
   g^{\alpha\beta}
   =
   \left(
    \begin{array}{rr}
     \frac43  & -\frac23 \\
     -\frac23 &  \frac43 \\
    \end{array}
   \right).
 \end{equation}
 The squared norm of, say, ${\bf M}$ is given by
 $\|{\bf M}\|^2=m_\alpha m^\alpha$ ($m^\alpha=g^{\alpha\beta}m_\beta$). 
 In \EQ{eq_L1}, the lattice constant of ${\cal R}^*$ is denoted as $a^*$
 ($=\sqrt{g^{11}}$).
 \setcounter{footnote}{0}
 \footnote{
 There are three 2D spaces:
 (i) The basal 2D space of $\Lambda$.
 Since the Cartesian components can be used for the position vector
 ${\bf x}$, we employ the alphabetical subscripts, $i$, $j$, to specify
 them.
 (ii) The 2D space in which ${\cal R}$ is embedded and (iii) its dual
 space in which ${\cal R}^*$ is embedded.
 Like the case of the crystallography, since we have employed the
 nonorthogonal primitive vectors, \EQ{eq_e^alpha}, for the 2D space to
 which ${\bf \Phi}$ belongs, it is necessary to introduce the dual space
 spanned by \EQ{eq_e_alpha}.
 In this case, it is convenient to express the vectors as the
 covariant/contravariant elements; we thus use the Greek alphabets,
 $\alpha$, $\beta$, $\gamma$, as the subscript/superscript, accordingly.
 }

 The action $S_0\equiv\int{\rm d}^2x\cL_0$ represents an interface model
 consisting of two kinds of massless scalar fields
 \cite{Kond95,Kond96}.
 Using the elements of ${\bf\Phi}$, it is rewritten as
 \begin{equation}
  S_0
   =
   \int {\rm d}^2x~
   \frac{K}{2\pi}\partial_i\phi_\alpha({\bf x})\partial_i\phi^\alpha({\bf x}),
   \label{eq_S0}
 \end{equation}
 where
 $\sqrt2\phi^\alpha\equiv{\bf e}^{\alpha}\cdot{\bf\Phi}$
 ($\phi_\alpha=g_{\alpha\beta}\phi^\beta$)
 [the element of ${\bf\Theta}$ is also defined by
 $\sqrt2\theta_\alpha\equiv{\bf e}_{\alpha}\cdot{\bf\Theta}$
 ($\theta^\alpha=g^{\alpha\beta}\theta_\beta$)].
 The factor $\sqrt2$ is for convenience.
 Then, the two-point function exhibits the logarithmic behavior
 \begin{equation}
  \AV{\phi^\alpha({\bf x})\phi^\beta({\bf 0})}{0}{}
   =
   -\frac{1}{4K}g^{\alpha\beta}\ln\left(\frac{r}{a}\right)^2,
   \label{eq_Green}
 \end{equation}
 where $r$ and $a$ are the distance between ${\bf x}$ and ${\bf 0}$ on
 the basal 2D space and the ultraviolet (UV) cutoff constant,
 respectively. 
 $\AV{\cdots}{0}{}$ means the average respect to the free part $S_0$.
 This implies that the fields themselves cannot represent physical
 quantities.
 However, as we summarize in appendix A, the current and the vertex
 operators defined by them are the scaling operators, and represent
 physical quantities.
 Since the system defined by $S_0$ is critical, and possesses the
 conformal invariance with $c=2$, the interface model is in a roughening
 phase if $\cL_{1,2}$ are both irrelevant.
 
 The phase locking potential $\cL_1$ consists of the six terms with the
 following electric vector charges [see \FIG{FIG2}(b)]:
 \begin{equation}
  \pm p {\bf e}^1,~
  \pm p {\bf e}^2,{\rm~~~and~~~}
  \pm p({\bf e}^1+{\bf e}^2),
  \label{eq_M}
 \end{equation}
 whose lengths are all $pa^*$.
 In the unit cell of ${\cal R}$, it produces the $p^2$ potential minima
 which form the triangular lattice as
 ${\bf\Phi}_{\rm lock}\equiv2\pi l^\alpha{\bf e}_\alpha/p$
 with
 $l^\alpha\in[0,p-1]$
 (see figure\ 1 in
 \cite{Otsu07}),
 and each of which corresponds to one of the $p^2$-degenerate states.
 From the formula, \EQ{eq_XMN}, the RG eigenvalue of $\cL_1$ is given by
 $2-2p^2/3K$ on the Gaussian fixed point $S_0$, so it becomes relevant
 for $K>p^2/3$.
 Since the Gaussian coupling $K$ stands for the stiffness of the
 interface, it is roughly proportional to the inverse temperature.
 Therefore, $\cL_1$ can stabilize the flat phase with the long-range
 order at low temperature.

 Another potential $\cL_2$ is defined in term of the dual field
 ${\bf \Theta}$.
 The vertex operator ${\rm e}^{{\rm i}{\bf N\cdot\Theta}}$ creates
 a discontinuity of ${\bf\Phi}$ by amount of $2\pi{\bf N}$ around the
 point ${\bf x}$.
 This topological defect is necessary to describe the disordered phase
 at high temperature.
 According to the RG sense, the most relevant terms are sufficient to be
 included in the potential, so the summation is performed only for the
 following magnetic vector charges with the shortest length 1
 [see \FIG{FIG2}(a)]:
 \begin{equation}
  \pm {\bf e}_1,~
  \pm {\bf e}_2,{\rm~~~and~~~}
  \pm({\bf e}_1-{\bf e}_2).
  \label{eq_N}
 \end{equation}
 From the formula \EQ{eq_XMN}, the RG eigenvalue of $\cL_2$ is given by
 $2-K/2$ on $S_0$.
 It thus becomes relevant for $K<4$ and brings about the disordered
 phase.

 Theoretically, it has been expected that the critical intermediate
 phase ($T_{\rm L}\le T\le T_{\rm H}$) survives for the case $p\ge4$
 \cite{Alca83},
 and also that the point $K=K_{\rm L}$ ($\equiv p^2/3$)
 [$K=K_{\rm H}$ ($\equiv4$)] where $\cL_1$ ($\cL_2$) becomes marginal
 corresponds to
 $T_{\rm L}$ ($T_{\rm H}$)
 \cite{Otsu07}.
 In
 \cite{Otsu07},
 we observed the existence of the critical intermediate phase for the
 $p=6$ case by the use of the numerical method.
 Further, since the effective field theory possesses the duality nature
 which is not obvious in the lattice model, we checked the validity of
 our theory on the self-dual point deep in the intermediate phase, and
 also estimated $T_{\rm L,H}$ semiquantitatively.
 Here, we shall perform the detailed analysis of the system in the
 vicinity of the two phase transition points.

 \figureII

 \subsection{Operator product expansions, three-point functions, and remarks}
 
 To utilize the CFT technology, we shall first clarify the relationship
 among local operators which plays an important role in our discussion.
 Since the intermediate region corresponds to the Gaussian fixed line
 parameterized by $K$, the so-called $\cM$ operator
 \cite{Kada79,Kada79B},
 \begin{equation}
  \cM({\bf x})
  \equiv
  \frac{Ka^2}{\sqrt8}
  \sum_{i=x,y}
  \|\partial_i{\bf\Phi}({\bf x})\|^2
  \label{eq_M_op}
 \end{equation}
 which is proportional to $\cL_0$ and translates the system along the
 line, is the most important one.
 The two-point function is given by
 $\AV{\cM({\bf x})\cM({\bf 0})}{0}{}=\left(a/r\right)^4$
 so that \EQ{eq_M_op} defines the truly marginal operator satisfying the
 normalization condition.
 Adding to this, we define the local operators proportional to
 $\cL_{1,2}$ as
 \begin{eqnarray}
 &\cV({\bf x})
  \equiv
  \frac{1}{\sqrt6}
  \sum_{\|{\bf M}\|=pa^*}:{\rm e^{i{\bf M\cdot\Phi  }({\bf x})}}:,
  \label{eq_V}\\
 &\cW({\bf x})
  \equiv
  \frac{1}{\sqrt6}
  \sum_{\|{\bf N}\|= 1  }:{\rm e^{i{\bf N\cdot\Theta}({\bf x})}}:.
  \label{eq_W}
 \end{eqnarray}
 Their two-point functions are 
 $\AV{\cV({\bf x})\cV({\bf 0})}{0}{}=\left({a}/{r}\right)^{2x_{\cV}}$
 and
 $\AV{\cW({\bf x})\cW({\bf 0})}{0}{}=\left({a}/{r}\right)^{2x_{\cW}}$
 with the dimensions
 \begin{equation}
  x_{\cV}\equiv\frac{2p^2}{3K}{\rm~~~and~~~}
  x_{\cW}\equiv\frac{K}{2}, 
 \end{equation}
 so they are also in the normalized forms.

 First, let us consider the expansion of the operator product
 $\cV({\bf x})\cV({\bf 0})$, which becomes important for
 $K\simeq K_{\rm L}$.
 While there are 36 terms in the double summations with respect to the
 vector charges, say, ${\bf M}$ and ${\bf M'}$, the following two cases
 are enough to be taken into account:
 (i) ${\bf M+M'=0}$ (six terms)
 and 
 (ii) $\|{\bf M+M'}\|=pa^*$ (12 terms); 
 the other 18 terms are irrelevant.
 After some calculations using the basic relations in appendices, we
 find that the cases (i) and (ii) mainly give $\cM$ and $\cV$,
 respectively.
 We then obtain the expression of the OPE as follows:
 \begin{equation}
  \cV({\bf x})\cV({\bf 0})
   \simeq
   -
   \frac{x_{\cV}}{\sqrt2}\left(\frac{a}{r}\right)^{2x_{\cV}-2}\cM({\bf 0})
   +
   \frac{2}{\sqrt6}\left(\frac{a}{r}\right)^{ x_{\cV}}\cV({\bf 0})
   +
   \cdots.
   \label{eq_OPE_VV}
 \end{equation}
 The part ``$\cdots$'' includes the unit operator, the stress tensor as
 well as less singular terms.
 It should be noted that the second term in the right-hand side (rhs)
 appears due to the
 triangular-lattice structure of ${\cal R}^*$, which is highly
 contrasted to the single component case, and brings about differences
 as we will see in the following.
 The cross-check of \EQ{eq_OPE_VV} can be done by performing the another
 OPE calculation
 \begin{equation}
  \cM({\bf x})\cV({\bf 0})
   \simeq
   -
   \frac{x_{\cV}}{\sqrt2}\left(\frac{a}{r}\right)^2\cV({\bf 0})
   +
   \cdots, 
   \label{eq_OPE_MV}
 \end{equation}
 which exhibits the symmetry property of the OPE coefficients to
 satisfy, i.e., $C_{\cal VVM}=C_{\cal MVV}$ ($=C_{\cal VMV})$.
 Now, we can read off the OPE coefficients as follows: 
 \begin{equation}
  C_{\cal VVM}=-\frac{x_{\cV}}{\sqrt2}{\rm~~~and~~~}
  C_{\cal VVV}=\frac{2}{\sqrt6}.
  \label{eq_OPEC_1}
 \end{equation}
 
 Next, we shall consider the region near $K\simeq K_{\rm H}$, and derive
 the OPE of $\cW({\bf x})\cW({\bf 0})$.
 Using the basic relations in appendices, it proceeds in parallel with
 the derivations of \EQS{eq_OPE_VV}{eq_OPE_MV}.
 Then, we obtain
 \begin{eqnarray}
 &\cW({\bf x})\cW({\bf 0})
  \simeq
  \frac{x_{\cW}}{\sqrt2}\left(\frac{a}{r}\right)^{2x_{\cW}-2}\cM({\bf 0})
  +
  \frac{2}{\sqrt6}\left(\frac{a}{r}\right)^{x_{\cW}}\cW({\bf 0})
  +
  \cdots,\\
 &\cM({\bf x})\cW({\bf 0})
  \simeq
  \frac{x_{\cW}}{\sqrt2}\left(\frac{a}{r}\right)^2\cW({\bf 0})
  +
  \cdots.
 \end{eqnarray}
 Thus, the OPE coefficients are given by
 \begin{equation}
  C_{\cal WWM}=\frac{x_{\cW}}{\sqrt2}{\rm~~~and~~~}
  C_{\cal WWW}=\frac{2}{\sqrt6},
  \label{eq_OPEC_2}
 \end{equation}
 where the nonzero $C_{\cal WWW}$ is again attributed to the triangular
 lattice structure of the repeat lattice ${\cal R}$.

 As one of the consequences of the above OPE calculations, we can fix
 the three-point functions among operators.
 In contrast to the single component case, we obtain the nonvanishing
 one for the phase locking potentials, e.g, 
 \begin{equation}
  \AV{\cV({\bf x}_1)\cV({\bf x}_2)\cV({\bf x}_3)}{0}{}
  =
  C_{\cal VVV}\prod_{1\le j<k\le3}\left(\frac{a}{r_{jk}}\right)^{x_{\cV}},
  \label{eq_THREE_VVV}
 \end{equation}
 where $r_{jk}$ is the distance between ${\bf x}_j$ and ${\bf x}_k$ 
 (the same relation also holds for $\cW$).
 This is because three vectors at the angle of 120 degrees to each other
 [e.g., $p {\bf e}^1$, $p{\bf e}^2$, and $-p({\bf e}^1+{\bf e}^2)$,
 as visible in \FIG{FIG2}(b)] satisfy
 {\it the vector charge neutrality condition}
 \cite{Alca83,Otsu07}
 (an extension of the scalar case
 \cite{Kada79,Kada79B}), 
 and this plays an important role in the following discussion.

 Lastly, we shall refer to other operators not listed above.
 The spin degrees of freedom is the most basic one, and is defined as
 $S_k\equiv:{\rm e}^{{\rm i}\varphi_k}$:.
 In the previous paper, based on its response to the spin rotations,
 \EQ{eq_gener},
 we argued that its sublattice dependent
 expression is given by
 \begin{equation}
 (S_{\rm a},S_{\rm b},S_{\rm c})
  =\left(
  :{\rm e}^{ {\rm i}({\bf e}^1+{\bf e}^2)\cdot{\bf \Phi}}:,
  :{\rm e}^{-{\rm i} {\bf e}^1           \cdot{\bf \Phi}}:,
  :{\rm e}^{-{\rm i} {\bf e}^2           \cdot{\bf \Phi}}:\right),
  \label{eq_SPIN}
 \end{equation}
 whose dimensions are all $x_S=2/3K$ [see \EQ{eq_XMN}]
 \cite{Otsu07}.
 This is one example of a general form of the quantities related to the
 spin degrees of freedom, i.e.,
 \begin{equation}
  {\cal O}({\bf x};\{w_{\bf M}\})
  \equiv
  \sum_{\|{\bf M}\|=M}w_{\bf M}:{\rm e}^{{\rm i}{\bf M}\cdot{\bf \Phi}}:~~~
  (w_{\bf M}\in{\mathbb C}). 
 \end{equation}
 For instance, for $S_{\rm a}({\bf x})$, the length of vector charges
 $M$ equals to $a^*$, and the weights are given by
 $(
 w_{ {\bf e}^1          },
 w_{ {\bf e}^1+{\bf e}^2},
 w_{ {\bf e}^2          },
 w_{-{\bf e}^1          },
 w_{-{\bf e}^1-{\bf e}^2},
 w_{-{\bf e}^2          })=(0,1,0,0,0,0)$.
 In the above, we observed that {\it the uniform mode} whose weights are
 independent of the direction of ${\bf M}$ is engaged in the Lagrangian
 density due to its symmetry property
 [see the definition of $\cV$ in \EQ{eq_V}].
 However, it is also expected that {\it the staggered mode} with
 $(
 w_{ p {\bf e}^1           },
 w_{ p({\bf e}^1+{\bf e}^2)},
 w_{ p {\bf e}^2           },
 w_{-p {\bf e}^1           },
 w_{-p({\bf e}^1+{\bf e}^2)},
 w_{-p {\bf e}^2           })=(+1,-1,+1,-1,+1,-1)$
 plays an important role, while we shall not discuss this issue in
 detail.

 \subsection{Renormalization-group equations}

 Since the data necessary for the use of the CFT technology have been
 obtained, we shall here perform the RG analysis of our effective field
 theory.
 First, we consider the region near $K\simeq K_{\rm L}$, where $\cL_2$
 is irrelevant.
 For convenience, we define the scaling field $y_0$ as
 \begin{equation}
  K=(1+y_0)K_{\rm L}.
 \end{equation}
 Then, the system can be described by the fixed-point Lagrangian density
 (or CFT) $\cL_0^{\rm L}$ (i.e., the Gaussian part with $K_{\rm L}$)
 perturbed by two marginal operators $\cM$ and $\cV$ as
 \begin{equation}
  \cL
  \simeq
  \cL_0+\cL_1
  =
  \cL_0^{\rm L}
  +
  \frac{\sqrt2y_0}{2\pi a^2}\cM({\bf x})
  +
  \frac{\sqrt6y_p}{2\pi a^2}\cV({\bf x}). 
  \label{eq_L_near_TL}
 \end{equation}
 For a perturbed CFT defined by the Lagrangian density,
 $\cL_{\rm gen.}=\cL_0^*+\sum_\mu\lambda_\mu{\cal O}_\mu({\bf x})/2\pi a^2$,
 where marginal scalar operators ${\cal O}_\mu$ are normalized as
 $\AV{{\cal O}_\mu({\bf x}){\cal O}_\nu({\bf 0})}{0}{*}
 =\delta_{\mu\nu}\left(a/r\right)^4$
 ($\AV{\cdots}{0}{*}$ means the average at the fixed point under
 consideration), the one-loop RG equations are governed by the OPE
 coefficients:
 For the change, $a\to(1+{\rm d}l)a$, they are given by
 ${\rm d}\lambda_\mu/{\rm d}l=
 -\frac12\sum_{\nu,\rho}C^*_{\mu\nu\rho}\lambda_\nu\lambda_\rho$
 ($C^*_{\mu\nu\rho}$ denotes the value on $\cL_0^*$)
 \cite{Poly72}. 
 In the present case, using coefficients of \EQ{eq_OPEC_1} at
 $K=K_{\rm L}$, we obtain the following equations:
 \begin{eqnarray}
  &\frac{{\rm d}y_0(l)}{{\rm d}l}=3y_p(l)^2,
  \label{eq_RG_TL_y0}\\
  &\frac{{\rm d}y_p(l)}{{\rm d}l}=2y_0(l)y_p(l)-y_p(l)^2.
  \label{eq_RG_TL_yp}
 \end{eqnarray}
 Similarly to the BKT RG-flow diagram, these exhibit one separatrix
 between the ordered and the critical phases, i.e.,
 \begin{equation}
  y_p(l)=-y_0(l), 
  \label{eq_SEP_TL}
 \end{equation}
 and one straight flow, $y_p(l)=2y_0(l)/3$, renormalized to the
 strong-coupling fixed point (see the right panel in \FIG{FIG3}).
 These are similar to those obtained in the research on the
 triangular-lattice defect melting problem
 \cite{Nels79,Houg80}
 (see also
 \cite{Alca83}). 
 Consequently, we can introduce the small parameter $t$ to control the
 distance from the separatrix as
 \begin{equation}
  y_p(l)=-(1+t)y_0(l)~~~(|t|\ll1).
  \label{eq_SEP_NTL}
 \end{equation}

 \figureIII

 Next, we shall derive the RG equations near $K\simeq K_{\rm H}$, where
 $\cL_1$ is irrelevant.
 We redefine the scaling field $y_0$ as
 \begin{equation}
  K=(1+y_0)K_{\rm H}.
 \end{equation}
 Then, the system is described by
 \begin{equation}
  \cL
  \simeq
  \cL_0+\cL_2
  =
  \cL_0^{\rm H}
  +
  \frac{\sqrt2y_0}{2\pi a^2}\cM({\bf x})
  +
  \frac{\sqrt6y_1}{2\pi a^2}\cW({\bf x}),
  \label{eq_L_near_TH}
 \end{equation}
 where $\cL_0^{\rm H}$ is the fixed-point Lagrangian density for the
 high-temperature transition.
 The RG equations are similarly obtained as follows:
 \begin{eqnarray}
  &\frac{{\rm d}y_0(l)}{{\rm d}l}=-3y_1(l)^2,
  \label{eq_RG_TH_y0}\\
  &\frac{{\rm d}y_1(l)}{{\rm d}l}=-2y_0(l)y_1(l)-y_1(l)^2.
  \label{eq_RG_TH_y1}
 \end{eqnarray}
 Since these are related to \EQS{eq_RG_TL_y0}{eq_RG_TL_yp} via the
 replacement
 $(y_0,y_p) \to (-y_0,y_1)$,
 one separatrix between the disordered and the critical phases,
 \begin{equation}
  y_1(l)=y_0(l),
  \label{eq_SEP_TH}
 \end{equation}
 and one straight flow, $y_1(l)=-2y_0(l)/3$, renormalized to the
 high-temperature fixed point, are embedded (see the left panel in
 \FIG{FIG3}).
 Thus, for the same aim, we shall introduce the small parameter $t$ as
 \begin{equation}
  y_1(l)=(1+t)y_0(l)~~~(|t|\ll1). 
  \label{eq_SEP_NTH}
 \end{equation}

 Here, we note that Boyanovsky and Holman performed the RG of the
 vectorial sine-Gordon field theory based on the simply-laced Lie
 algebras
 \cite{Boya91}.
 While they provided a general argument on properties of operators,
 and some of them are the same as those observed above, we have focused
 on $\cL$ corresponding to the specific model, TSIM.

 \subsection{Mixing of marginal operators}\label{ss_HMO}

 According to one of the present authors' discussion for the sine-Gordon
 field theory, linear combinations of marginal operators play
 an important role
 \cite{Nomu95}.
 As we see in the following, it is true also in the present case. 
 So, we shall consider this issue in this subsection. 
 Let us start with the system around the separatrix \EQ{eq_SEP_TL}, and
 consider the following quantities:
 \begin{equation}
  \cA\propto\cM+c_1\cV{\rm~~~and~~~}
  \cB\propto\cV+c_2\cM.
 \end{equation}
 The two real coefficients $c_{1,2}$ are to be determined from the
 orthogonality condition
 $\AV{\cA({\bf x}_1)\cB({\bf x}_2)}{}{}=0$
 which persists under the renormalization along the separatrix [while
 the normalization conditions, e.g.,
 $\AV{\cA({\bf x}_1)\cA({\bf x}_2)}{}{}=\left({a}/{r_{12}}\right)^4$,
 are used to determine overall constants].
 Instead of the correlation function, we consider a more convenient
 quantity
 \cite{Giam89}:
 \begin{equation}
  F(r_{12},a,y_0(l),y_p(l))
  \equiv
  \left(\frac{r_{12}}{a}\right)^4\AV{\cA({\bf x}_1)\cB({\bf x}_2)}{}{},
 \end{equation}
 and evaluate $F$ and its response to the change of the cutoff
 ${\rm d}F/{\rm d}l$ up to the lowest order in the coupling constants 
 $y_0$ and $y_p$.
 For this, we first expand $F$ with respect to $y_p$ as
 \begin{equation}
  F(r_{12},a,y_0,y_p)
  \simeq
  F_0(r_{12},a,y_0)-\sqrt6y_pF_1(r_{12},a,y_0), 
  \label{eq_F_perturb}
 \end{equation}
 where
 \begin{eqnarray}
  F_0
  &=
  \left(\frac{r_{12}}{a}\right)^4
  \AV{\cA({\bf x}_1)\cB({\bf x}_2)}{0}{},
  \label{eq_F_0}\\
  F_1
  &=
  \left(\frac{r_{12}}{a}\right)^4
  \int\frac{{\rm d}^2x_3}{2\pi a^2}
  \AV{\cA({\bf x}_1)\cB({\bf x}_2)\cV({\bf x}_3)}{0}{}. 
  \label{eq_F_1}
 \end{eqnarray}
 We should regularize the UV divergence of the integral over
 ${\bf x}_3$ in \EQ{eq_F_1} by excluding two circles of the radius $a$
 centered at ${\bf x}_1$ and ${\bf x}_2$.
 Explicitly, the integral is restricted as
 \begin{equation}
  \int\to\int H(r_{13}-a)H(r_{23}-a),
 \end{equation}
 where $H(x)$ is the Heaviside step function.
 Noticing $4-2x_{\cV}\simeq4y_0$, we can rewrite \EQ{eq_F_0} as
 $F_0\simeq c_1+c_2+4c_1y_0 \ln\left(r_{12}/a\right)$.
 This exhibits $F$ being almost constant $F\simeq c_1+c_2$, so the
 condition in the lowest order,
 \begin{equation}
  c_1+c_2=0,
  \label{eq_C_0}
 \end{equation} 
 should be satisfied.
 Next, let us consider its response to the change of the cutoff,
 ${\rm d}F_0/{\rm d}l$.
 There exist two types of contributions, i.e., 
 (i) a direct one via the cutoff $a$
 and
 (ii) an indirect one via the coupling constant $y_0$
 controlled by the RG equations.
 Since the \GES{$\beta$-functions}{eq_RG_TL_y0}{eq_RG_TL_yp} only
 include the second-order terms, we can neglect the latter.
 Then, we obtain
 \begin{equation}
  \frac{{\rm d}F_0}{{\rm d}l}\simeq -4c_1y_0.
  \label{eq_dF_0/dl}
 \end{equation}
 The contributions from the change of the coupling constant $y_p$ to the
 response of the second term in \EQ{eq_F_perturb} can be neglected due
 to the same reason, so we shall consider
 ${\rm d}F_1/{\rm d}l$ up to the zeroth order in $y_0$.
 Like the case of the first term, there also exist two types of
 contributions; we can omit the type (ii) contributions.
 Furthermore, as we have already seen in the derivation of
 \EQ{eq_dF_0/dl}, a part of the type (i) contributions stemming from the
 power-of-$a$ factors and giving the $\Or(y_0)$ terms can be neglected.
 Consequently, the response is contributed only from the change of $a$
 in the UV regularization factor
 \begin{equation}
  \fl~~~~~~~~~~
  \frac{{\rm d}F_1}{{\rm d}l}
  \simeq
  \left(\frac{r_{12}}{a}\right)^4
  \int\frac{{\rm d}^2x_3}{2\pi a^2}~
  \AV{\cA({\bf x}_1)\cB({\bf x}_2)\cV({\bf x}_3)}{0}{\rm L}~
  \frac{{\rm d}}{{\rm d}l}\left[H(r_{31}-a)H(r_{32}-a)\right].
  \label{eq_dF_0/dl_1}
 \end{equation}
 Since the integral is a line one along two circumferences of circles
 centered at ${\bf x}_1$ and ${\bf x}_2$, we can estimate the
 rhs of \EQ{eq_dF_0/dl_1} by using the asymptotic form of the
 three-point functions [e.g., \EQ{eq_THREE_VVV}];
 the result is the following: 
 \begin{equation}
  \frac{{\rm d}F_1}{{\rm d}l}
  \simeq
  -2\left[C^{\rm L}_{\cal MVV}(1+c_1c_2)+C_{\cal VVV}c_1\right],
  \label{eq_dF_0/dl_2}
 \end{equation}
 where $C^{\rm L}_{\cal MVV}=-\sqrt2$.
 Consequently, from the lowest-order calculation of the condition
 ${\rm d}F/{\rm d}l=0$,
 we obtain the relation
 $c_1y_0+\left[\sqrt3(1+c_1c_2)-c_1\right]y_p=0$.
 On the separatrix $y_p=-y_0$, this is reduced to
 \begin{equation}
  2c_1-\sqrt3(1+c_1c_2)=0,
  \label{eq_C_1}
 \end{equation}
 which, together with \EQ{eq_C_0}, can determine the coefficients. 
 While the quadratic equation for $c_1$, $c_1^2+2c_1/\sqrt3-1=0$,
 possesses two solutions $1/\sqrt3$ and $-\sqrt3$, both of these provide
 an identical description of the operators.
 Thus, in the following discussion, we choose $c_1=1/\sqrt3$, and call
 $\cA$ and $\cB$ as the $\cM$-like and the $\cV$-like operators,
 respectively
 (see reference \cite{Nomu95}).
 Their normalized expressions are then given by
 \begin{equation}
  \left(
   \begin{array}{c}
    \cA \\
    \cB
   \end{array}
  \right)
  =
  \left(
   \begin{array}{cc}
    ~~\cos\vartheta_{\rm L}&\sin\vartheta_{\rm L} \\
     -\sin\vartheta_{\rm L}&\cos\vartheta_{\rm L}
   \end{array}
  \right)
  \left(
   \begin{array}{c}
    \cM \\
    \cV
   \end{array}
  \right)
 \end{equation}
 with $\tan\vartheta_{\rm L}=1/\sqrt3$.
 Here, note the followings:
 Since the condition to determine the mixing angle $\vartheta_{\rm L}$,
 \EQ{eq_C_1}, is expressed in terms of the OPE coefficients, we can
 recognize it as an appearance of the universal properties of the fixed
 point $\cL_0^{\rm L}$.
 Further, in the single component case, the corresponding mixing angle
 is given by $\tan\vartheta=1/\sqrt2$
 \cite{Nomu95}.
 This difference mainly stems from the $C_{\cal VVV}$ contribution
 absent in the scalar case.

 Next, let us move on to the region near $T_{\rm H}$, and consider the
 system around the separatrix \EQ{eq_SEP_TH}, where the following linear
 combinations are to be determined:
 \begin{equation}
  \cC\propto\cM+d_1\cW{\rm~~~and~~~}
  \cD\propto\cW+d_2\cM.
 \end{equation}
 In the same way as the above, the orthogonality condition
 $\AV{\cC({\bf x}_1)\cD({\bf x}_2)}{}{}=0$
 persisting under the renormalization along the separatrix determines
 the real coefficients $d_{1,2}$.
 Since the calculations are performed in parallel with the above case,
 we can straightforwardly derive the equations corresponding to
 \EQS{eq_C_0}{eq_C_1} as
 $d_1+d_2=0$ and $2d_1+\sqrt3(1+d_1d_2)=0$,
 respectively.
 In accordance with the above case, the solution $d_1=-1/\sqrt3$ is
 chosen, so that $\cC$ and $\cD$ are termed as the $\cM$-like and the
 $\cW$-like operators, respectively (the difference between $\cA$ and
 $\cC$ is contextually obvious).
 The normalized expressions are then given as follows:
 \begin{equation}
  \left(
   \begin{array}{c}
    \cC \\
    \cD
   \end{array}
  \right)
  =
  \left(
   \begin{array}{cc}
    ~~\cos\vartheta_{\rm H}&\sin\vartheta_{\rm H}\\
     -\sin\vartheta_{\rm H}&\cos\vartheta_{\rm H}
   \end{array}
  \right)
  \left(
   \begin{array}{c}
    \cM \\
    \cW
   \end{array}
  \right)
 \end{equation}
 with $\tan\vartheta_{\rm H}=-1/\sqrt3$.
 Consequently, independently of $p$, we find a simple relation between
 the mixing angles at the high- and the low-temperature transitions,
 $\vartheta_{\rm L}=-\vartheta_{\rm H}$ $(=\pi/6)$. 

 \subsection{Corrections to finite-size scaling and eigenvalue structures}

 We are in position to calculate the renormalized scaling dimensions of
 operators around the fixed points $\cL_0^{\rm L,H}$ and to discuss the
 significance of the results above.
 We shall start from the free part defined on an infinitely long
 cylinder in the $x_2$ direction with a periodicity of $L$ in the
 $x_1$ direction, and write the partition function using the action
 $S_{0,\rm cyl.}\equiv\int_{\rm cyl.}{\rm d}^2x\cL_0$ as
 $Z_{0,\rm cyl.}
  \equiv
  \int[{\rm d}{\bf \Phi}]~{\rm e}^{-S_{0,\rm cyl.}}
  \propto
  \lim_{\tau\to\infty}{\rm Tr}~{\rm e}^{-\tau\hat H_{0,L}}$.
 Then, $\hat H_{0,L}$ exhibits the Hamiltonian operator associated with
 the transfer matrix ${\rm e}^{-\hat H_{0,L}}$; it defines a 1D quantum
 system with length $L$, and is given by 
 $\hat H_{0,L}=\int_0^L{\rm d}x_1\hat\cH_0(x_1)$ with
 \begin{equation}
  \hat\cH_0(x)
  =
  \frac{v}{2}
  \left[
  \frac{\pi}{K} \hat \pi_\alpha(x) \hat \pi^\alpha(x)
  +
  \frac{K}{\pi} \partial_x \hat \phi_\alpha(x) \partial_x \hat \phi^\alpha(x)
  \right]. 
 \end{equation}
 The momentum $\hat \pi_\alpha$ conjugate to the field operator
 $\hat\phi^\alpha$ satisfies
 $[\hat\phi^\alpha(x),\hat\pi_\beta(x')]
 ={\rm i}\delta^\alpha_\beta\delta(x-x')$,
 and $v$ is the velocity of an elementary excitation.
 When we writing its eigenvalue and eigenstate as
 $E_{0,L,\nu}$ and $\bigl|\nu\rangle$,
 the CFT provides the finite-size-scaling form of the excitation gap as 
 $\Delta E_{0,L,\nu}\equiv E_{0,L,\nu}-E_{0,L,{\rm g}}=2\pi vx_\nu/L$,
 where $E_{0,L,{\rm g}}$ and $x_\nu$ are the lowest energy and the
 scaling dimension of the operator corresponding to the state
 $\bigl|\nu\rangle$ ($L$ is supposed to be large enough).
 Next, we consider the Hamiltonian density corresponding to the generic
 model $\cL_{\rm gen.}$, i.e., 
 $\hat\cH_{\rm gen.}(x)=
 \hat\cH_0^*(x)+\sum_\mu\lambda_\mu\hat{\cal O}_\mu(x)/2\pi a^2$.
 Writing the ground-state and the excited-state energies as
 $E_{L,{\rm g}}$ and $E_{L,\nu}$,
 then we can calculate the corrections to scaling within the first-order
 perturbation as
 $\Delta E_{L,\nu}
  \equiv
  E_{L,\nu}-E_{L,{\rm g}}
  \simeq
  \frac{2\pi v}{L}(x_\nu+\sum_\mu\lambda_\mu C^*_{\mu\nu\nu})$
 \cite{Card86}.
 The parenthesized quantity in the rhs defines the renormalized scaling
 dimension. 
 Using the OPE coefficients, the mixing angle $\vartheta_{\rm L}$, and
 this formula, we obtain the dimensions of the $\cM$-like and the
 $\cV$-like operators near $T_{\rm L}$ [i.e., near the separatrix
 \EQ{eq_SEP_NTL}] as
 \begin{eqnarray}
  &x_{\cA}=2+2y_0\left(1+\frac54t\right), \label{eq_xA}\\
  &x_{\cB}=2-6y_0\left(1+\frac34t\right).
 \end{eqnarray}
 Similarly, we obtain those of the $\cM$-like and the $\cW$-like
 operators near $T_{\rm H}$ [\EQ{eq_SEP_NTH}] as
 \begin{eqnarray}
  &x_{\cC}=2-2y_0\left(1+\frac54t\right),\label{eq_xC}\\
  &x_{\cD}=2+6y_0\left(1+\frac34t\right)
 \end{eqnarray}
 ($y_0$ was redefined as mentioned above).
 Since these corrections to scaling are described by the OPE
 coefficients, there are some universal relations among the dimensions.
 For instance, in the present case, we find that
 \begin{eqnarray}
  \frac{3x_{\cA}+x_{\cB}}{4}=2& {\rm~~on~~}y_p=-y_0,
  \label{eq_TL_av}\\
  \frac{3x_{\cC}+x_{\cD}}{4}=2& {\rm~~on~~}y_1= y_0.
  \label{eq_TH_av}
 \end{eqnarray}
 Since the ratio of the level splitting, 1:3, being different from that
 in the single component case
 \cite{Nomu95}
 is one of features,
 this can provide a solid evidence of the
 BKT-like phase transition described by the RG
 \GES{equations}{eq_RG_TL_y0}{eq_RG_TL_yp} or
 \GES{equations}{eq_RG_TH_y0}{eq_RG_TH_y1}.

 \section{Numerical calculations (the $p=6$ case)} \label{sec_NUMERICAL}

 In this section, we shall explain our numerical calculations and
 results to confirm the above theoretical predictions.
 We consider the system on $\Lambda$ with $M$ ($\to\infty$) rows of $L$
 (a multiple of 3) sites wrapped on the cylinder, and define the
 transfer matrix connecting the next-nearest-neighbor rows
 (see \FIG{FIG1}).
 We denote its eigenvalues as $\lambda_q(L)$ or their logarithms as
 $E_q(L)=-\frac12\ln|\lambda_q(L)|$ ($q$ specifies a level).
 Then, the conformal invariance in critical systems provides the
 expressions of the central charge $c$ and the scaling dimension $x_q$
 as the corrections to scaling
 \cite{Card84,Blot86,Affl86}:
 \begin{equation}
  E_{\rm g}(L)\simeq Lf-\frac{\pi}{6L\zeta}c{\rm~~~and~~~}
  \Delta E_q(L)\simeq \frac{2\pi}{L\zeta}x_q. 
 \end{equation}
 Here, $E_{\rm g}(L)$, $\Delta E_q(L)$ $[=E_q(L)-E_{\rm g}(L)]$, $\zeta$
 $(=2/\sqrt3)$, and $f$ correspond to ``the ground-state energy'', ``the
 excitation gap'', the geometric factor, and a free energy density,
 respectively.
 In numerical diagonalization calculations using the Lanczos algorithm,
 we employ two fundamental spin rotations $\hat R_{\rm a}$
 and $\hat R_{\rm b}$ in \EQ{eq_gener} as well as the lattice
 translation and the space inversion.
 This is because the matrix size can be reduced, and more importantly
 discrete symmetries can specify lower-energy excitations.
 For instance, since the spin degrees of freedom on $\Lambda_{\rm a}$
 transforms as
 $\hat R_{\rm a}S_k\mapsto {\rm e}^{{\rm i}2\pi/p}S_k$ and 
 $\hat R_{\rm b}S_k\mapsto S_k$, 
 the corresponding excitation level can be found in the sector with
 indexes (${\rm e}^{{\rm i}2\pi/p}$,1), and provides a small scaling
 dimension $x_S=2/3K$
 \cite{Otsu07}.
 Thus, we shall utilize also this level for the determinations of the
 phase transition points.

 \figureIV

 First, we consider the system around the separatrix \EQ{eq_SEP_TL}.
 From \EQ{eq_xA} and the dimension of the sublattice dependent spin,
 $x_S\simeq 2(1-y_0)/p^2$, the condition
 \begin{equation}
  p^2x_S=4-x_{\cA}
  \label{eq_LC_TL}
 \end{equation}
 is satisfied at $t=0$ [i.e., on the separatrix \EQ{eq_SEP_TL}].
 Thus, it can be employed as a criterion to determine the
 low-temperature transition point $T_{\rm L}$.
 We perform the numerical calculations for the $p=6$ case and for the
 systems up to $L=9$.
 In \FIG{FIG4}(a), we exhibit temperature dependences of the scaling
 dimensions, i.e., the both sides of \EQ{eq_LC_TL} estimated for the
 $L=9$ site system, and find the level crossing at which the condition
 is satisfied.
 Therefore, we obtain the finite-size estimate, $1/T_{\rm L}(L)$, from
 the crossing point.
 As we show in the inset, the extrapolation of finite-size estimates
 to the thermodynamic limit is performed based on the
 least-squares-fitting procedure
 $1/T_{\rm L}(L)=1/T_{\rm L}+a/L^2$. 
 Then, we obtain the transition point as $1/T_{\rm L}\simeq\inverseTL$.

 Second, we consider the determination of $T_{\rm H}$, which can be
 performed in the parallel way to the above.
 From \EQ{eq_xC} and the dimension of spin, $x_S\simeq (1-y_0)/6$,
 around \EQ{eq_SEP_TH}, the condition
 \begin{equation}
  12x_S=x_{\cC}
  \label{eq_LC_TH}
 \end{equation}
 is found to be satisfied only on the separatrix.
 In \FIG{FIG4}(b), we provide the same plot as \FIG{FIG4}(a), where the
 circles (squares) with the fitting curve plot the rhs (lhs) of
 \EQ{eq_LC_TH}.
 Again, we find the level crossing at which the condition is satisfied.
 Therefore, we can estimate $1/T_{\rm H}(L)$ from the crossing point.
 The extrapolation to the limit $L\to\infty$ is also performed (see the
 inset), and then we estimate the high-temperature transition point as
 $1/T_{\rm H}\simeq\inverseTH$.
 
 In the previous paper, we roughly estimated the transition points from
 the behavior of the central charge (i.e., deviations from the
 theoretical value $c=2$), and obtained $1/T_{\rm L,H}\simeq1.5$, 1.1,
 respectively. 
 Thus, our above estimates through the level crossings are found to be
 consistent with the data of the central charge.
 
 \figureV

 At this stage, it is important to check the universal relations among
 the scaling dimensions.
 As mentioned, since the relation, \EQ{eq_TL_av}, must hold between the
 $\cM$-like and the $\cV$-like excitations at $T_{\rm L}$, we calculate
 the average (i.e., its lhs) at $1/T=1.51$.
 As is shown in \FIG{FIG5}(a), the estimates converge to the theoretical
 value 2 very accurately (see the circles with fitting line), meanwhile
 the scaling dimensions themselves considerably deviate from 2 (see the
 up- and the down-ward triangles).
 Further, the relation $(3p^2x_S-x_{\cB})/2=2$ is also expected to hold
 at the transition point, so we calculate the difference (the $p=6$
 case), and plot the data in the same figure (see the square with the
 fitting line).
 Despite the smallness of the system sizes, the relation holds within
 5\% error.
 These checks can be passed only if the system is at the BKT-like phase
 transition point, and the numerically utilized levels possess the
 theoretically expected properties.
 Therefore, these are helpful to demonstrate the reliability of our
 approach and results.

 We perform the same checks for the high-temperature transition.
 In \FIG{FIG5}(b), we plot the average [the lhs of \EQ{eq_TH_av}] at
 $1/T=1.05$.
 We find the excellent convergence of the data to 2 in the thermodynamic
 limit (see the circles with fitting line).
 In addition, another relation $(36x_S+x_{\cD})/4=2$ is expected between
 the dimensions of the spin and the $\cW$-like operators; we plot the
 average in the same figure (see the square with fitting line).
 Then, we find the deviation of the limiting value ($\simeq 2.13$) from
 2.
 This may be due to the following reason: 
 In the thermodynamic limit, the universal jump of $K(l=\infty)$ from
 $K_{\rm H}$ (=4) to 0 occurs at $T_{\rm H}$, and $x_S$ is inversely
 depending upon $K$ [see \EQ{eq_XMN}].
 Therefore, $x_S$ is sensitive to the temperature.
 $T_{\rm H}$ was reliably estimated from the level crossing, but due to
 the limitation in the size of systems treated, it may include some
 error which causes the deviation.
 
 Consequently, we have applied the level-crossing conditions to
 determine the BKT-like transition points, and then we have checked some
 universal relations among excitation levels at the transition points.
 This strategy (the level spectroscopy) was proposed and developed by
 one of the present authors to analyze the BKT transitions in 1D quantum
 spin systems
 \cite{Nomu95}.
 In that case, the sine-Gordon field theory is relevant to the
 discussion.
 On the other hand, since the present BKT-like transitions are described
 by the vector sine-Gordon models, we have extended the strategy to be
 applicable to them.
 Then, we have successfully demonstrated its efficiency through the
 numerical calculations of TSIM.

 \section{Discussion and summary} \label{sec_DISCUSSION}

 Up to now, we have concentrated on the properties of the critical
 intermediate phase:
 It possesses the conformal symmetry with $c=2$, and exhibits the
 transitions to the ordered and the disordered phases.
 While, for the latter, its similarity to the triangular-lattice
 defect melting phenomena was argued in the literature
 \cite{Alca83,Otsu07},
 we shall re-visit the universality class of the transition, and refer
 to its relevance to a ground-state phase transition observed in a
 quantum spin chain system.
 
 In the limit $p\to\infty$, the symmetry of TSIM
 (${\mathbb Z}_p\times{\mathbb Z}_p$) becomes the U(1)$\times$U(1),
 i.e., the continuous one, and eliminates the low-temperature ordered
 phase.
 The RG-flow \GES{equations}{eq_RG_TH_y0}{eq_RG_TH_y1} still describe
 the transition to the disordered phase, and enable us to analyze the
 system around the transition point.
 Based on them,  quite recently, one of the authors proposed a
 finite-size-scaling ansatz for the helicity modulus of TSIM
 \cite{Otsu08},
 which has a great relevance to the transition
 (see also \cite{Fish73,Nels77,Ohta79}).
 The ansatz reflects a self-similarity of trajectories with respect to a
 conserved quantity of the flow, and mainly predicts the following:
 (i)
 In the disordered phase, the correlation length is given by
 $\xi\propto\exp[{\rm const}/(T-T_{\rm H})^{\bar{\nu}}]$
 with the exponent $\bar{\nu}=3/5$. 
 (ii)
 The finite-size-scaling function takes a universal value at the
 transition temperature, which comes from the RG flow along the
 separatrix $y_1=y_0$.
 While we performed large-scale Monte-Carlo (MC) simulations of TSIM in
 this limit to verify the predictions, here we only mention that
 simulation data exhibit a good agreement with the ansatz (the readers
 interested in the detailed discussions and the calculation results may
 refer to reference \cite{Otsu08}).
 Since it includes the prediction also in the disordered phase, its
 confirmation is complementary to the present argument of
 \SES{sec_THEORY}{sec_NUMERICAL},
 and thus provides another solid evidence to support our theoretical
 description.
 Simultaneously,
 the exponent $\bar{\nu}=3/5$ does not agree with the previous research
 \cite{Alca83},
 where $\bar{\nu}=2/5$ was predicted based on the vector CG
 representation and the RG argument on the triangular-lattice defect
 melting theory
 \cite{Youn79}. 
 Also, in the previous paper\
 \cite{Otsu07},
 the explanation that $\bar\nu$ takes 2/5 was given based on their
 arguments.
 But, now we are confident that the exponent should be 3/5, so the
 reason of this discrepancy should be clarified in future.

 Instead of in the 2D classical systems, we can find the same situation
 in the ground state of a 1D quantum system.
 TSIM is invariant under the symmetry group $S_3$ of the sublattice
 permutations, and exhibits the conformal invariance with $c=2$ which
 stems from the continuous symmetry in the $p\to\infty$ limit.
 Guided by these properties, we are led to think of the quantum lattice
 gas model with three components because it realizes the $S_3$ symmetry
 as permutations of components, and its exact solution shows the $c=2$
 criticality
 \cite{Uimi70,Lai74,Suth75}.
 Also, as its extension, the bilinear-biquadratic (BLBQ) spin-1 chain is
 widely known, and is defined by the Hamiltonian:
 \begin{equation}
  H_{\rm BLBQ}
  =
  \sum_{\langle j,k\rangle}
  \left[
  \cos\theta     ~{\bf S}_j\cdot{\bf S}_k+
  \sin\theta\left({\bf S}_j\cdot{\bf S}_k\right)^2
  \right].
 \end{equation}
 This model possesses some points where the exact information is
 available:
 The Affleck-Kennedy-Lieb-Tasaki point
 \cite{Affl87}, 
 the Takhatajan-Babujian (TB) point ($\theta_{\rm BT}=-\pi/4$)
 \cite{Takh82,Babu82},
 and
 the Uimin-Lai-Sutherland (ULS) point ($\theta_{\rm ULS}=\pi/4$)
 \cite{Uimi70,Lai74,Suth75}.
 The last one which corresponds to the quantum lattice gas model
 separates the extended critical phase
 ($\pi/2\ge\theta\ge\theta_{\rm ULS}$)
 \cite{Fath91,Fath93}
 and the Haldane phase
 ($\theta_{\rm ULS}>\theta>\theta_{\rm TB}$)
 \cite{Hald83},
 and it is described by the level-1 SU(3) Wess-Zumino-Witten model.
 The central charge for the former and the correlation length in the
 latter were calculated as $c=2$ and
 $\xi\propto\exp\left[{\rm const}/(\theta_{\rm ULS}-\theta)^{3/5}\right]$,
 respectively
 \cite{Itoi97},
 with which the numerical estimations agree
 \cite{Fath91,Fath93}.
 According to the analysis by Itoi and Kato, the critical fixed line
 does not exist around the ULS point, so the global RG-flow diagram is
 considerably different from the present one
 \cite{PC0}.
 However, the transition occurs when the system crosses the separatrix
 with the SU(3) symmetry, and if we focusing on the massive region
 including the transition point, the RG feature is seemingly similar to
 our case.
 This may be a reason why the exponent $\bar{\nu}$ takes the same value
 in both cases.
 Furthermore, quite recently, the SU($N$) self-dual sine-Gordon model
 consisting of the $(N-1)$-component vectorial fields has been
 investigated
 \cite{Lech06,Lech07};
 its relevances to, for instance, the quantum-spin chains (including the
 BLBQ model) and ladders have been indeed clarified
 \cite{Lech06}.
 For TSIM, we have seen that $\cL_0$ consists of the two current
 operators $j^{\alpha}(z)$ $(\alpha=1,2)$.
 In addition, for instance for $K=K_{\rm H}$, the potential $\cL_2$
 becomes marginal, and the vector charges \EQ{eq_N} in ${\cal R}$ which
 is isomorphic to the root lattice of the SU(3) Lie algebra provide the
 six operators $v_{\bf K}(z)$ with the conformal weight
 $(\Delta,\overline{\Delta})=(1,0)$ (see appendix B). 
 Thus, there exist eight chiral current operators, and they may define
 the level-1 SU(3) current algebra
 \cite{Fren80,Sega81},
 as in the case of the Kagom\'e-lattice three-state Potts
 antiferromagnet
 \cite{Kond95,Kond96}.
 From these all,
 we think that although TSIM is in the lower symmetry than that of the
 BLBQ model, it exhibits a symmetry enhancement at the end points of the
 intermediate phase, and then it may share the same fixed point
 properties with the ULS model while more concrete evidences are
 desired.

 Lastly, we shall comment on the $p=4$ case.
 Although our theory in \SEC{sec_THEORY} as well as the vector CG
 analysis
 \cite{Alca83}
 predicts the intermediate phase for $4\le K\le16/3$, the previous MC
 data indicated a sign of the first-order transition between
 the ordered and the disordered phases
 \cite{Alca82}.
 However, subsequent studies in computational physics revealed that it
 is in general difficult to distinguish between the weak first-order and
 the second-order transitions just based on MC data
 \cite{PeczakLandau89,LeeKosterlitz90}. 
 Further, there are considerable difficulties also in the treatment of
 the BKT-like phase transitions (e.g., an accurate determination of the
 transition point) by MC methods.
 Thus, for the $p=4$ case, the nature of phase transitions does not seem
 to be established yet.

 As one of the possibilities, other than \EQM{eq_L0}{eq_L1}{eq_L2},
 there may exist a term inferred from the symmetry consideration, and it
 might eliminate the intermediate phase as speculated before [see
 figure\ 7(b) in
 \cite{Alca83}].
 Nevertheless, we think that this issue remains as an important future
 problem.

 To summarize, we have investigated the BKT-like continuous phase
 transitions observed in the triangular-lattice three-spin interaction
 model (TSIM) based on the vector dual sine-Gordon field theory.
 The basic properties of the local density operators (e.g., the scaling
 dimensions) and their mutual relations (the OPE coefficients) have been
 investigated in detail.
 Using these CFT data, we have performed the RG analysis of phase
 transitions and the conformal perturbation calculations of the
 excitation spectra up to the one-loop order.
 Especially, the mixing angles of the marginal operators on the
 separatrixes for the low-temperature and the high-temperature
 transitions, i.e., $\vartheta_{\rm L,H}$, have been determined and
 compared to the single component case.
 Then, we have found some universal relations among the renormalized
 scaling dimensions, which can precisely characterize the present phase
 transitions.
 Furthermore, we have pointed out their importance for the numerical
 determinations of the phase transition points.
 To check the theory, we performed the numerical diagonalization
 calculations of the transfer matrix of TSIM (the $p=6$ case) up to the
 system size $L=9$, and determined the transition points as $1/T_{\rm
 L,H}=\inverseTL,\inverseTH$, respectively, which was followed by the
 check of the universal relations among the excitation levels.
 Lastly, we have discussed the enhancement of symmetry at the end points
 of the critical intermediate phase.
 Based on the existence of the eight current operators and the value of
 the exponent $\bar\nu=3/5$, we have argued its relevance to the ground
 state of the bilinear-biquadratic spin-1 chain.

 \ack

 We thank
 S Hayakawa, M Iihosi, H Matsuo, and M Fujimoto
 for stimulating discussions. 
 We would like to also thank
 P Lecheminant
 for drawing our attention to the references\ \cite{Boya91,Lech06,Lech07}.
 Main computations were performed using the facilities of
 Information Synergy Center in Tohoku University,
 Cybermedia Center in Osaka University,
 and
 Yukawa Institute for Theoretical Physics
 in Kyoto University.
 This work was supported by
 Grants-in-Aid from the Japan Society for the Promotion of Science,
 Scientific Research (C),
 No\ 17540360 and No\ 18540376.

 \appendix

 \section{Two-dimensional massless scalars: the operator product
 expansions and the conformal invariance}
 
 The action in \EQ{eq_S0} consists of two massless scalars located in the
 2D Euclidean space.
 Here, we summarize its basic properties, e.g., the equation of motion,
 the operator product expansions, and the conformal invariance
 \cite{Polchinski}.

 It is convenient to adopt the complex coordinates
 $z, \bar z=x\pm{\rm i}y$ (the former takes the upper sign).
 When we define
 $  \phi^\alpha(z,\bar z)\equiv  \phi^\alpha({\bf x})$,
 $\theta_\alpha(z,\bar z)\equiv\theta_\alpha({\bf x})$,
 ${\rm d}^2z\equiv2{\rm d}^2x$,
 and
 $\partial, \bar\partial\equiv(\partial_x\mp{\rm i}\partial_y)/2$,
 then \EQ{eq_S0} is expressed as
 \begin{equation}
  S_0
  =
  \int {\rm d}^2z~
  \frac{K}{\pi}
  \partial\phi_\alpha(z,\bar z)
  \bar\partial\phi^\alpha(z,\bar z). 
 \end{equation}
 The classical equation of motion is then
 \begin{equation}
  \partial\bar\partial\phi^\alpha(z,\bar z)=0,
 \end{equation}
 which exhibits the chiral decomposition of fields, i.e.,
 \begin{equation}
  \phi^\alpha(z, \bar z),~
  \theta^\alpha(z, \bar z)
  =\frac{K^{\mp\frac12}}{2}[\psi^\alpha(z)\pm\bar\psi^\alpha(\bar z)].
 \end{equation}
 In terms of new fields with only holomorphic or antiholomorphic
 dependence, the action is re-expressed as
 \begin{equation}
  S_0
  =
  \int {\rm d}^2z~
  \frac{1}{4\pi}
  \partial\psi_\alpha(z)
  \bar\partial\bar\psi^\alpha(\bar z),
 \end{equation}
 and their two-point functions are diagonal in the sense that
 \begin{eqnarray}
  \AV{\psi^\alpha(z)\psi^\beta(0)}{0}{}
  =
  -g^{\alpha\beta}\ln\frac{z}{a},
  \label{eq_two-point-z}\\
  \AV{\bar\psi^\alpha(\bar z)\bar\psi^\beta(0)}{0}{}
  =
  -g^{\alpha\beta}\ln\frac{\bar z}{a},
  \label{eq_two-point-z-bar}
 \end{eqnarray}
 which otherwise vanish.
 These show that $\psi^\alpha$ and $\bar\psi^\alpha$ are not the scaling
 operators.
 However, their derivatives
 \begin{equation}
      j^\alpha(     z)
 \equiv{\rm i}a     \partial    \psi^\alpha(     z){\rm~~~and~~~}
 \bar j^\alpha(\bar z)
 \equiv{\rm i}a \bar\partial\bar\psi^\alpha(\bar z)
 \label{eq_j}
 \end{equation}
 exhibit, e.g.,
 $\AV{j^\alpha(z)j^\beta(0)}{0}{}=g^{\alpha\beta}\left(a/z\right)^2$,
 so that
 $j^\alpha$ and $\bar j^\alpha$
 are candidates of those with the scaling dimension 1.
 As usual, this issue can be confirmed by the OPE with the stress tensor
 which is obtained by the Noether theorem:
 It is diagonal in the complex coordinates, and is given by
 \begin{equation}
           T (     z )
 =\frac12:     j_\alpha(     z )     j^\alpha(     z ):{\rm~~~and~~~}
 \overline{T}(\bar{z})
 =\frac12:\bar j_\alpha(\bar{z})\bar j^\alpha(\bar{z}):.
 \end{equation}
 Using the Wick theorem and the Taylor expanding, the OPE's can be
 obtained as follows:
 \begin{eqnarray}
 &T(z)j^\alpha(0)
  \simeq
  \left(\frac{a}{z}\right)^2j^\alpha(0)
  +
  \left(\frac{a}{z}\right)^1a\partial j^\alpha(0), \\
 &\overline{T}(\bar z)\bar{j}^\alpha(0)
  \simeq
  \left(\frac{a}{\bar z}\right)^2\bar{j}^\alpha(0)
  +
  \left(\frac{a}{\bar z}\right)^1a\bar\partial\bar{j}^\alpha(0). 
 \end{eqnarray}
 These exhibit that $j^\alpha$ and $\bar{j}^\alpha$ are the scaling
 operators with the conformal weights
 $(\Delta,\overline{\Delta})=(1,0)$ and $(0,1)$,
 respectively.
 The vertex operators are also important examples of the scaling
 operators; they are introduced by
 \begin{equation}
       v_{     \bf k }(     z)
  \equiv:{\rm e}^{{\rm i}     k_\alpha    \psi^\alpha(     z)}:{\rm~~~and~~~}
  \bar v_{\bar{\bf k}}(\bar z)
  \equiv:{\rm e}^{{\rm i}\bar k_\alpha\bar\psi^\alpha(\bar z)}:.
  \label{eq_v}
 \end{equation}
 The two-point functions behave as, e.g.,
 $\AV{v_{\bf k}(z)v_{-\bf k}(0)}{0}{}=\left(a/z\right)^{\|{\bf k}\|^2}$,
 where $k_\alpha$ is the covariant element of a constant vector ${\bf k}$
 and $\|{\bf k}\|^2\equiv k_\alpha k^\alpha$.
 Similarly to the above, the OPE's of $Tv$ and $\overline{T}\bar v$ are
 given as follows:
 \begin{eqnarray}
 &T(z)v_{\bf k}(0)
  \simeq
  \frac{\|{\bf k}\|^2}{2}\left(\frac{a}{z}\right)^2v_{\bf k}(0)
  +
  \left(\frac{a}{z}\right)^1a\partial v_{\bf k}(0),
  \label{eq_Tv}\\
 &\overline{T}(\bar z)\bar{v}_{\bar {\bf k}}(0)
  \simeq
  \frac{\|\bar{\bf k}\|^2}{2}
  \left(\frac{a}{\bar z}\right)^2\bar{v}_{\bar {\bf k}}(0)
  +
  \left(\frac{a}{\bar z}\right)^1a\bar\partial\bar{v}_{\bar {\bf k}}(0). 
  \label{eq_Tvbar}
 \end{eqnarray}
 Thus, $v_{\bf k}$ and $\bar{v}_{\bar {\bf k}}$ are the scaling
 operators with the weights
 $(\|{\bf k}\|^2/2,0)$ and $(0,\|\bar{\bf k}\|^2/2)$,
 respectively.

 As we have seen, although the physical quantities possess both the
 holomorphic and the antiholomorphic parts, the OPE's are performed
 independently in these two parts due to the diagonal nature of the
 two-point \GES{functions}{eq_two-point-z}{eq_two-point-z-bar}.
 Therefore, for a while, we focus only on the holomorphic part.
 The OPE of $T$ with itself is given by
 \begin{eqnarray}
  T(z)T(0)
  \simeq
  \frac{\delta^\alpha_\alpha}{2}\left(\frac{a}{z}\right)^4
  +
  2\left(\frac{a}{z}\right)^2 T(0)
  +
  \left(\frac{a}{z}\right)^1a\partial T(0). 
 \end{eqnarray}
 Thus, we can read off the central charge as $c=\delta^\alpha_\alpha=2$,
 which is equal to the number of components of the vector field.
 The OPE between $j_\alpha$ and $v_{\bf k}$ are given by
 \begin{equation}
  j_\alpha(z)v_{\bf k}(0)
  \simeq
  k_\alpha\left(\frac{a}{z}\right)^1v_{\bf k}(0).
  \label{eq_jv}
 \end{equation}
 This indicates that $j_\alpha$ is the current operator to detect the
 $\alpha$th element of the vector charge $\bf k$ in the vertex operator.
 Further, the OPE between two vertex operators plays a very important
 role in our discussion; it can be expressed in the following form:
 \begin{equation}
  v_{\bf k}(z)v_{\bf -k'}(0)
  \simeq
  \left(\frac{a}{z}\right)^{\bf k\cdot k'}
  :v_{\bf k-k'}(0)\left[1+\Or\Bigl(\frac{z}{a}\Bigr)\right]:. 
 \end{equation}
 For the case ${\bf k\ne k'}$, we can neglect the $\Or\left(z/a\right)$
 terms in the rhs. 
 However, for ${\bf k=k'}$, since $v_{\bf 0}(z)=\hat{1}$ by definition,
 they become important.
 By expansion, we find
 \begin{equation}
  \frac{z}{a}
  k_\alpha j^\alpha(0)
  +
  \frac12\left(\frac{z}{a}\right)^2 
  \left\{k_\alpha a\partial j^\alpha(0)+
  \left[k_\alpha j^\alpha(0)\right]^2\right\},
 \end{equation}
 where the $\Or\left((z/a)^3\right)$ terms are dropped.

 \section{Some useful relations}

 In this appendix, we shall derive some useful relations which will be
 referred to in the discussion of \SEC{sec_THEORY}.
 Using \EQ{eq_j}, the $\cM$ operator in \EQ{eq_M_op} is given by
 \begin{equation}
  \cM({\bf x})=-\frac{1}{\sqrt2}j_\alpha(z)\bar j^\alpha(\bar z). 
 \end{equation}
 On the other hand, using \EQ{eq_v}, the vertex operator with the vector
 charges, $\bf M$ and $\bf N$, is expressed as
 \begin{equation}
  :{\rm e}^{{\rm i}\left[{\bf M\cdot\Phi}({\bf x})+{\bf N\cdot\Theta}({\bf x})\right]}:
  =
  v_{\bf K}(z)\bar v_{\overline{\bf K}}(\bar z)
 \end{equation}
 with
 ${\bf K},~\overline{\bf K}\equiv(K^{-\frac12}{\bf M}\mp K^{+\frac12}{\bf N})/\sqrt{2}$.
 From \EQS{eq_Tv}{eq_Tvbar}, we can obtain the formula for the scaling
 dimension of the vertex operator,
 $x_{\bf M,N}=\frac12\left(\|{\bf K}\|^2+\|\overline{\bf K}\|^2\right)$.
 It is rewritten as a function of the vectors:
 \begin{equation}
  x_{\bf M,N}=\frac12(K^{-1}\|{\bf M}\|^2+K\|{\bf N}\|^2). 
  \label{eq_XMN}
 \end{equation}
 The OPE between the $\cM$ operator and the vertex operator is
 calculated by using \EQ{eq_jv} as
 \begin{equation}
  \cM({\bf x})
  :{\rm e}^{{\rm i}\left[{\bf M\cdot\Phi}({\bf 0})+{\bf N\cdot\Theta}({\bf 0})\right]}:
  \simeq
  -\frac{{\bf K}\cdot\overline{\bf K}}{\sqrt2}
  \left|\frac{a}{z}\right|^2
  :{\rm e}^{{\rm i}\left[{\bf M\cdot\Phi}({\bf 0})+{\bf N\cdot\Theta}({\bf 0})\right]}:,
 \end{equation}
 where the coefficient is also given by
 ${\bf K}\cdot\overline{\bf K}=x_{\bf M,0}-x_{\bf 0,N}$.

 The OPE's between the vertex operators with opposite vector charges are
 the most important part in our calculations.
 Here, we consider the following quantity:
 \begin{equation}
  {\cal Q}
  \equiv
  \frac16\sum_{\|{\bf M}\|=pa^*}
  :{\rm e}^{ {\rm i}{\bf M\cdot\Phi}({\bf x})}:
  :{\rm e}^{-{\rm i}{\bf M\cdot\Phi}({\bf 0})}:,
 \end{equation}
 where the summation is over the six vectors in \EQ{eq_M}.
 The product of the holomorphic and the antiholomorphic parts gives many
 terms.
 Among them, those of the first order in the elements of the vector
 charge ${\bf M}$ disappear after the summation.
 For the second-order terms, by utilizing the relation,
 \begin{equation}
  \frac16\sum_{\|{\bf M}\|=pa^*}m_\alpha m_\beta=\frac{2p^2}{3}g_{\alpha\beta},
  \label{eq_MMg}
 \end{equation}
 we find the following compact expression:
 \begin{equation}
  {\cal Q}
  \simeq
  \left|\frac{a}{z}\right|^{4p^2/3K}
  \Bigl\{
  1
  +
  \frac{p^2}{3K}
  \Bigl[
  \left(\frac{     z}{a}\right)^2           T (0)
  +
  \left(\frac{\bar z}{a}\right)^2 \overline{T}(0)
  -
  \left|\frac{z}{a}\right|^2 \sqrt2\cM({\bf 0})
  \Bigr]
  \Bigr\}.
 \end{equation}
 Similarly, we can perform the OPE calculation of the following quantity
 \begin{equation}
  {\cal R}
  \equiv
  \frac16\sum_{\|{\bf N}\|=1}
  :{\rm e}^{ {\rm i}{\bf N\cdot\Theta}({\bf x})}:
  :{\rm e}^{-{\rm i}{\bf N\cdot\Theta}({\bf 0})}:,
 \end{equation}
 where the summation is over the six vectors given in \EQ{eq_N}.
 Like \EQ{eq_MMg}, the relation between the elements of the vector
 charge ${\bf N}$ and the metric tensor,
 \begin{equation}
  \frac16\sum_{\|{\bf N}\|=1}n^\alpha n^\beta=\frac12g^{\alpha\beta},
 \end{equation}
 is available.
 So, one can find the expansion
 \begin{equation}
  {\cal R}
   \simeq
   \left|\frac{a}{z}\right|^K
   \Bigl\{
   1
   +
   \frac{K}{4}
   \Bigl[
   \left(\frac{     z}{a}\right)^2           T (0)
   +
   \left(\frac{\bar z}{a}\right)^2 \overline{T}(0)
   +
   \left|\frac{z}{a}\right|^2 \sqrt2\cM({\bf 0})
   \Bigr]
   \Bigr\}.
 \end{equation}
 Consequently, we see that the OPE's include the secondary operators
 $T$ with $(\Delta,\overline{\Delta})=(2,0)$ and $\overline T$ with
 (0,2) as well as the $\cM$ operator with $(1,1)$
 (see also \cite{Kadanoff}).
 In general, a rotationally invariant system defined on the plain does
 not include $T$ and $\overline T$ because they possess the conformal
 spins with length 2.
 On the other hand, since the $\cM$ operator is scalar and proportional
 to $\cL_0$, these results naturally exhibit the renormalizations of the
 Gaussian coupling $K$ caused by the potentials $\cL_{1,2}$, and may
 also indicate that the metric tensor has been properly employed to
 define the fixed-point Lagrangian density $\cL_0$.

 \newcommand{\AxS}[1]{#1}
 \newcommand{\AxD}[2]{#1 and #2}
 \newcommand{\AxT}[3]{#1, #2 and #3}
 \newcommand{\AxQ}[4]{#1, #2, #3 and #4}
 \newcommand{\REF }[4]{#4 {\it #1} {\bf #2} #3}
 \newcommand{\PRA }[3]{\REF{Phys. Rev.\  {\rm A}}{#1}{#2}{#3}}
 \newcommand{\PRB }[3]{\REF{Phys. Rev.\  {\rm B}}{#1}{#2}{#3}}
 \newcommand{\PRE }[3]{\REF{Phys. Rev.\  {\rm E}}{#1}{#2}{#3}}
 \newcommand{\NPB }[3]{\REF{Nucl. Phys.\ {\rm B}}{#1}{#2}{#3}}

\end{document}